\documentclass{article}
\usepackage[utf8]{inputenc}
\usepackage[lmargin=0.65in,rmargin=0.65in,tmargin=0.9in,bmargin=0.75in]{geometry}
\usepackage{hyperref,lmodern,textcomp,graphicx}
\usepackage[table]{xcolor} \usepackage{caption,fancyhdr}
\usepackage{color,amsmath,amssymb,mathpazo,mathtools}
\usepackage{vwcol}
\usepackage{lipsum}
\usepackage{tgpagella}
\usepackage{indentfirst}
\usepackage{bchart}
\usepackage{lettrine,multicol}
\usepackage[sorting=none,style=chem-angew,autocite= superscript]{biblatex}
\usepackage{tcolorbox}%
\definecolor{mycolor}{rgb}{0.122, 0.435, 0.698}
\makeatletter
\newcommand{\mybox}[1]{%
  \setbox0=\hbox{#1}%
  \setlength{\@tempdima}{\dimexpr\wd0+13pt}%
  \begin{tcolorbox}[boxrule=0pt,arc=0pt,
      left=6pt,right=6pt,top=6pt,bottom=6pt,boxsep=1pt,width=\textwidth]
    #1
  \end{tcolorbox}
}

\fancypagestyle{firstpage}
{\fancyhead[L]{\textsc {\color {gray} {\large }}}    
\fancyhead[R]{\large \textsc{\color {gray} {Letter}}}}
\newcommand{\HA}[1]{{\color{black} #1}}
\graphicspath{{Figures/}}

\newcommand{\HBA}[1]{{\color{black} #1}}

\begin{document}

\thispagestyle{firstpage}
{\noindent \LARGE \textit{Brillouin-zone definition in non-reciprocal Willis monatomic lattices}} \\ [0.5em] 

\noindent {\large \textit{Hasan B. Al Ba'ba'a}} \\

\noindent \begin{tabular}{c >{\arraybackslash}m{6in}}
    \includegraphics[]{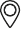} &
    \noindent {Department of Mechanical Engineering, Union College, Schenectady, NY 12308, USA} \\[0.5em]
    \includegraphics[]{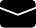}& \noindent \text{\href{mailto:albabaah@union.edu}{albabaah@union.edu}}\\
    
\end{tabular} \\


\mybox{\textit{Brillouin-zone~(BZ)~definition in a class of non-reciprocal Willis monatomic lattices (WMLs) is analytically quantified.~It is shown that BZ boundaries \textit{only} shift in response to non-reciprocity in one-dimensional WMLs, implying a \textit{constant} BZ width, with asymmetric dispersion diagrams exhibiting unequal wavenumber ranges for forward and backward going waves.~An extension to square WMLs is briefly discussed, analogously demonstrating the emergence of shifted and irregularly shaped BZs, which maintain \textit{constant} areas regardless of non-reciprocity strength.
}}

\vspace{0.25cm}
\noindent \textbf{Keywords
}

\noindent Non-reciprocal monatomic lattices; Brillouin-zone shift; Phase shift; Willis materials; Wave equation.

\noindent\rule{7.2in}{0.5pt}

\section{Introduction}
Exploration of new wave phenomena has been a driving motivation in wave propagation research \cite{wang2020tunable}, with a growing interest in elastodynamic non-reciprocity \cite{nassar2020nonreciprocity}.~\HA{Within linear, time-invariant elastic media}, a wave emanating from a source towards a receiver must not behave differently if the \HA{locations} of the source and receiver are interchanged, thanks to elastodynamic reciprocity.~Designs of artificial materials have enabled new avenues to \HA{manipulate} waves beyond what is possible in natural materials, and breakage of wave reciprocity is no exception. \HA{Given its potential in enabling novel applications (e.g.,~one-way vibrational isolators \cite{trainiti2016non})}, researchers have introduced a variety of structural designs \HA{tailored for non-reciprocity}, such as exploiting nonlinear instability with supra-transmission phenomenon~\cite{wu2019wave}. \HA{Comparably, granular crystals, with their nonlinear and bifurcating-chaotic behavior, have been exploited to asymmetrically transmit energy, with potential application in sensing and energy harvesting~\cite{boechler2011bifurcation}}.~\HA{Another} choice to achieve non-reciprocity is by spatiotemporally modulating mechanical properties, \HA{with artificial momentum bias forcing waves to propagate differently in the forward and backward directions}~\cite{trainiti2016non,nassar2017non,nassar2017modulated,attarzadeh2018non,attarzadeh2018wave,nassar2018quantization}.

\HA{When elastic media defy elastodynamic reciprocity, their dispersion diagram becomes skewed and asymmetric about a zero wavenumber, and conventional Brillouin-zone (BZ) definitions are rendered obsolete. Cassedy and Oliner are amongst the first contributors to study the skewed (non-reciprocal) nature of BZ in an attempt to generalize its definition for wave propagation in space-time modulated electrical circuits~\cite{cassedy1963dispersion}. Unlike spatial-only modulation with dispersion diagrams perfectly repeating along the wavenumber axis, the presence of time modulation creates dispersion-diagram periodicity along a line with a slope that is a function of such modulation (under small modulation amplitude assumption). However, the adequacy of traditional BZs to fully capture wave non-reciprocity remains an ongoing concern, especially for two-dimensional systems \cite{attarzadeh2018non}. 

In this effort, BZ definition is unraveled for a class of non-reciprocal monatomic lattices (MLs), synthesized from a non-reciprocal wave equation governing longitudinal waves in an axially-moving elastic (continuum) rod as follows \cite{attarzadeh2018elastic}:}
\begin{equation}
    \rho \frac{\partial^2 u}{\partial t^2} + 2\rho v_0 \frac{\partial^2 u}{\partial x \partial t} +\left(\rho v^2_0- E \right) \frac{\partial^2 u}{\partial x^2} = 0
    \label{eq:EOM-moving-rod}
\end{equation}
where $\rho$, $E$, and $u(x,t)$ denote the density, modulus of elasticity, and the longitudinal displacement as a function of space $x$ and time $t$ of the rod, respectively. The rod's \textit{constant} speed is defined as $v_0 = \beta c$, where $c = \sqrt{E/\rho}$ is the elastic-medium's sonic speed and \HA{$\beta \in [-1,1]$ ($+$ ($-$) sign indicates forward (backward) motion)} is the non-dimensional relative moving velocity~\cite{attarzadeh2018elastic}. \HA{As a matter of fact, the nature of non-reciprocity of Eq.~(\ref{eq:EOM-moving-rod}) is intertwined with that of elastic rods with spatiotemporally modulated properties.~At the low-frequency limit and slow modulation, the macroscopic behavior of an elastic rod with a spatiotemporal wave-like modulation is governed by a Willis-type equation~\cite{nassar2017modulated}, which is \textit{identical} in form to that of Eq.~(\ref{eq:EOM-moving-rod}) for moving rods~\cite{attarzadeh2018elastic}. Not only that Eq.~(\ref{eq:EOM-moving-rod}) is of Willis type, but it is also gyroscopic as it can be re-casted as~\cite{wickert1990classical}:
\begin{equation}
    M \frac{\partial^2 u}{\partial t^2} + G \frac{\partial u}{\partial t} + K u = 0
    \label{eq:EOM_operators}
\end{equation}
where $M = \rho$, $G = 2 \rho v_0 \partial_x$, and $K = (\rho v_0^2 - E) \partial^2_x $ are the mass, gyral, and stiffness operators.~Whether subjected to spatiotemporal modulation or motion at a constant speed, elastic rods obeying Eq.~(\ref{eq:EOM-moving-rod}) exhibit linear momentum bias (inducing non-reciprocity), as evident from the mixed derivative term.~The ramifications of the induced non-reciprocity on BZ shall be first established for the lattice version of Eq.~(\ref{eq:EOM-moving-rod}) and then extended to a two-dimensional counterpart.}

\begin{figure*}[]
     \centering
\includegraphics[]{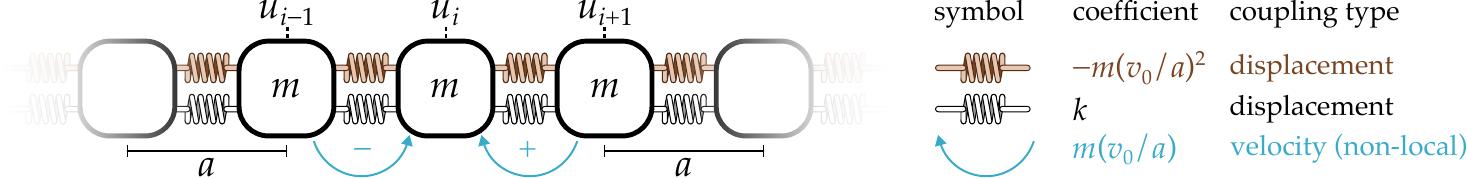}
\caption{Schematic of a non-reciprocal \HA{Willis} monatomic lattice (WML), \HA{synthesized} from discretizing Eq.~(\ref{eq:EOM-moving-rod}).~The lattice's masses $m$ are evenly spaced with a lattice constant $a$ and the displacement of the $i^\text{th}$ mass is donated as $u_i$. As seen from the figure, two types of springs exist: (i) a conventional spring $k$ arising from the elasticity of the rod and (ii) a negative spring $-m(v_0/a)^2$ induced as a consequence of the rod's motion. Non-local coupling parameters that are proportional to the velocity of the next neighbors of an $i^\text{th}$ unit cell break the lattice's reciprocity (shown for the $i^\text{th}$ unit cell only).}
     \label{fig:Sch}
\end{figure*}

\section{\HA{One-dimensional Willis} monatomic lattices}
\subsection{Mathematical model}
To find an equivalent ML to that of the moving rod, henceforth referred to as \textit{Willis monatomic lattices} (WMLs), the spatial derivatives in Eq.~(\ref{eq:EOM-moving-rod}) are discretized in the spatial domain $x$ using the central differencing scheme, resulting in
\begin{equation}
\rho \ddot{u}_{i} + \rho v_0 \left(\frac{\dot{u}_{i+1} - \dot{u}_{i-1}}{a}\right) + \left(\rho v_0^2 - E \right) \left(\frac{{u}_{i+1} - 2u_i + u_{i-1} }{a^2} \right)  = 0
\label{eq:step2}
\end{equation}
where $a$ is the finite difference spacing and constitutes the lattice's constant (Figure~\ref{fig:Sch}).~The spring constant and mass of WML are defined as the effective stiffness $k = EA/a$ and mass $m = \rho Aa$ of a rod segment of length $a$ and area $A$. Using the introduced parameters, a few mathematical manipulations result in the equation of motion for the $i^\text{th}$ unit cell of \HA{WML}:
\begin{equation}
    m \ddot{u}_{i} + \frac{m v_0}{a} \left( \dot{u}_{i+1} - \dot{u}_{i-1} \right) + \left(k -\frac{m v_0^2}{{a^2}} \right)\left(2u_i - {u}_{i+1}-u_{i-1} \right) = 0
    \label{eq:EOM_moving_monatomic_lattice}
\end{equation}
\HA{Examining Eq.~(\ref{eq:EOM_moving_monatomic_lattice}) and Figure~\ref{fig:Sch}, it is first deduced that the motion of individual masses $m$ is physically coupled through springs $k$, as in typical MLs \cite{Hussein2014,albabaa2018TDF}.  Having the momentum bias introduced by the rod's constant speed, the physics of the lattice's motion is influenced in two ways:}
\begin{enumerate}
    \item \HA{Besides the spring $k$, the $i^\text{th}$ unit-cell displacement is coupled with its next neighbors through negative springs of coupling coefficient $-mv_0^2/a^2$, which remains negative regardless of the sign of $v_0$ (or $\beta$).~Consequently, the effective stiffness ($k_e = k - m v_0^2/{a^2}$) of the WML reduces as the rod's speed $v_0$ increases, jeopardizing dynamical stability when $m v_0^2/{a^2}> k$ (equivalent to $|\beta|>1$) as a consequence of its effective stiffness $k_e$ becoming negative.} 
    \item Non-local coupling terms emerge, which are proportional to the velocity of next-neighboring masses of the $i^{\text{th}}$ unit cell.~Such non-local interactions (\HA{related to} Willis coupling \cite{nassar2022waves}) arise from discretizing the mixed derivative in Eq.~(\ref{eq:EOM-moving-rod}) and dictate the strength of lattice's non-reciprocity. While the physical motion of the elastic medium enables such non-local coupling, it may be alternatively achieved via feedback control~\cite{Rosa2020DynamicsInteractions,braghini2021non,pechac2021non}. \HA{Lastly, it is noteworthy that such non-local (positive-negative) couplings are akin to skew-symmetric gyroscopic coupling studied in literature \cite{carta2014dispersion,Nash2015TopologicalMetamaterials,wang2015topological,susstrunk2016classification,attarzadeh2019non,zhou2023simple}.}
\end{enumerate}

\subsection{Dispersion relation}
Assuming harmonic motion $u_i = \hat{u}_i \text{e}^{ \textbf{i} \omega t}$ and applying Bloch boundary conditions $u_{i\pm 1} = \text{e}^{\pm \textbf{i} q} u_i$, \HA{where $q$ is the non-dimensional wavenumber, $\omega$ is the excitation frequency, and $\mathbf{i}$ is the imaginary unit}, a non-dimensional dispersion relation is derived from Eq.~(\ref{eq:EOM_moving_monatomic_lattice}):
\begin{equation}
      \Omega^2 + 2 \Omega \beta \sin(q) - 4(1-\beta^2)\sin^2 \left( \frac{q}{2} \right) = 0
    \label{eq:non-dim_disp_rel}
\end{equation}
Here, the definition of the non-dimensional frequency is $\Omega = \omega/\omega_0$, where $\omega_0 = \sqrt{k/m}$. Note that $v_0 = \beta c \equiv a \beta \omega_0$. The dispersion relation in Eq.~(\ref{eq:non-dim_disp_rel}) has a single (positive-frequency) dispersion \HA{branch}, which reads:
\begin{equation}
    \Omega = 2\left|\sin\left(\frac{q}{2}\right)\right|\sqrt{1-\beta^2 \sin^2\left(\frac{q}{2}\right)} - \beta \sin(q)
\label{eq:disp_branch}
\end{equation} \HA{The term $\sin(q)$ is what makes the dispersion relation in (\ref{eq:disp_branch}) non-reciprocal as its sign changes simultaneously with $q$.} Corroborating our earlier observation regarding the lattice's dynamical stability with the speed~$v_0$, Eq.~(\ref{eq:disp_branch}) indicates that $|\beta| >1$ yields complex values of the frequency $\Omega$, signaling dynamical instability.

\begin{figure*}[hb!]
     \centering
\includegraphics[]{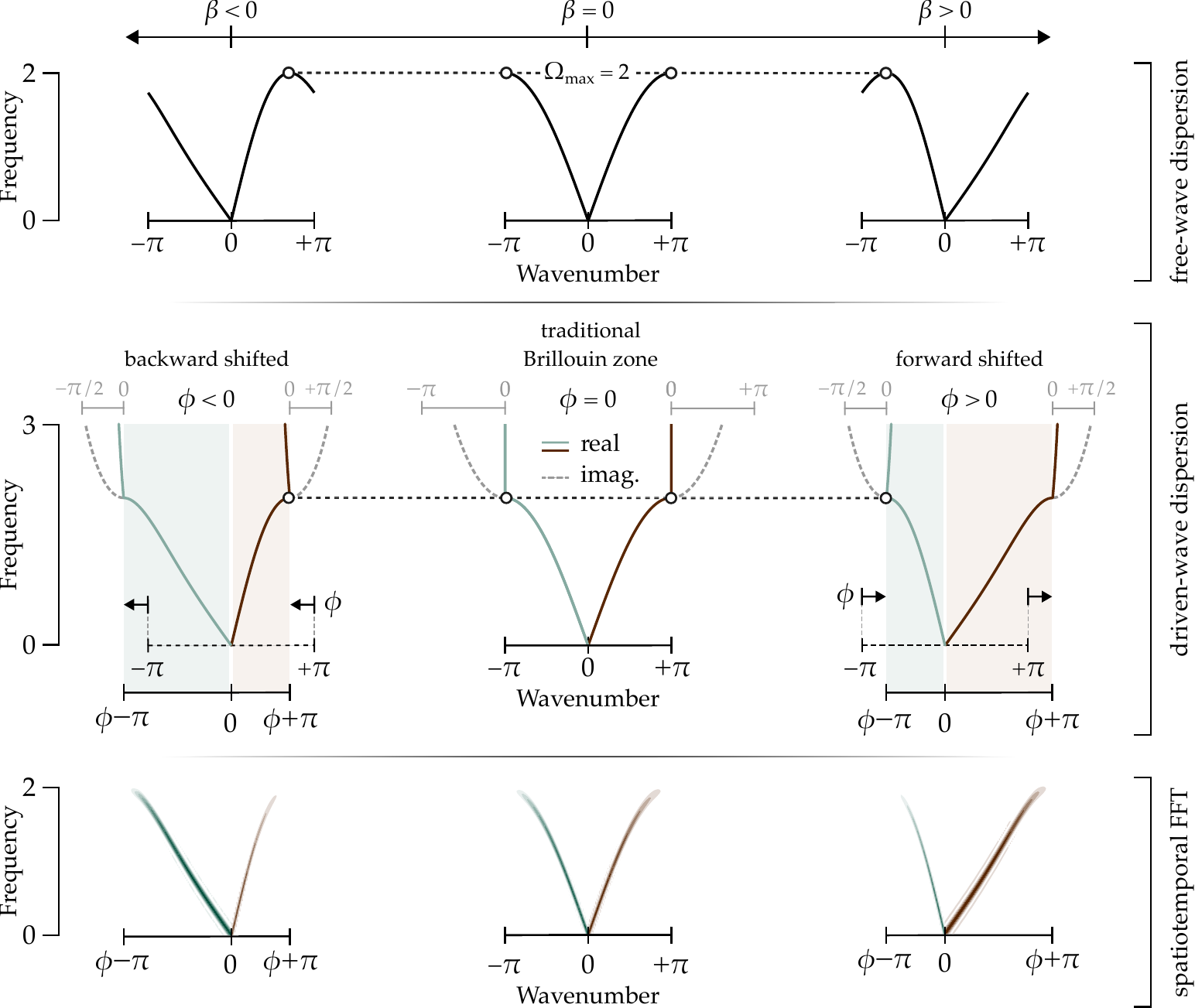}
\caption{\HA{Dispersion diagram of WMLs, depicted using free-wave (top) and driven-wave (middle) methodologies}.~A positive (negative) $\beta$ results in a positive (negative) shift $\phi$, as seen from the bias \HA{towards} the forward-going (backward-going) waves, \HA{while $\beta=0$ recovers the dispersion diagram of a reciprocal ML}.~However, the free-wave dispersion does not capture the shifted nature of BZ, showing the driven-wave approach significance in unraveling the BZ shift $\phi$. \HA{Observe} that the attenuation (represented by the \HA{imaginary} component of the wavenumber \HA{in the driven-wave dispersion}) is smaller in \HA{WMLs relative to its reciprocal ML counterpart ($\beta=0$) when $\Omega>2$, i.e., a frequency higher than the lattice's cutoff frequency $\Omega_{\text{max}}=2$}.~Also, $\Omega>2$ results in a non-constant real component of the wavenumber in WML cases only.~\HA{Note that, for $\beta = \pm 0.5$ shown here, the corresponding BZ shift is $\phi \approx 0.3\pi$.} \HA{(bottom) Spatiotemporal FFT for the time response of WML for an impulse excitation in the left and right ends, quantifying forward- and backward-going wavenumbers, respectively, and demonstrating excellent agreement with the new BZ definition in the middle panel.}}
     \label{fig:Dispersion}
\end{figure*}

\subsection{Brillouin-zone definition}

Conventionally, the dispersion relation \HA{is} plotted by sweeping the non-dimensional wavenumber within the first BZ of the range $q\in[-\pi,\pi]$ and solve for $\Omega$ in Eq.~(\ref{eq:disp_branch}) for each value of $q$, a method known as \textit{free-wave} dispersion relation. The results of this process are shown in the top row of Figure~\ref{fig:Dispersion} for different values of $\beta$, where $\beta \neq 0$ induces dispersion asymmetry about zero wavenumber (i.e., $q = 0$). Note that $\beta = 0$ recovers the dispersion diagram of a classical (and reciprocal) ML \cite{albabaa2018TDF} and results in a perfectly symmetric dispersion about the center of BZ ($q = 0$). However, the conventional BZ does not accurately capture \HA{the range of wavenumbers spanning the forward-going} and backward-going waves for $\beta \neq 0$.~To \HA{show} that, Eq. (\ref{eq:non-dim_disp_rel}) can be reformulated to solve for the wavenumber $q$ by having the frequency $\Omega$ as an input, resulting in the \textit{driven-wave} dispersion relation.~After rewriting the trigonometric terms in their exponential form, one can show that:
\begin{equation}
     (1-\beta^2 - \textbf{i} \beta \Omega) e^{\textbf{i}q} + \left(\Omega^2 - 2(1-\beta^2) \right) + (1-\beta^2 + \textbf{i} \beta \Omega)e^{-\textbf{i}q} = 0
    \label{eq:non-dim_driven_wave}
\end{equation}
Observe that the first and last terms in Eq.~(\ref{eq:non-dim_driven_wave}) are complex conjugates, which can be rewritten as $|z| e^{\pm\textbf{i}(q-q_s)}$, where
\begin{equation}
    |z| = \sqrt{\beta^2 \Omega^2 + (1-\beta^2)^2}; \ \ \ \ \
    q_s = \tan^{-1}\left(\frac{\Omega \beta}{1-\beta^2} \right)
    \label{eq:mag_phase}
\end{equation}
\HA{The emergent} \textit{phase shift} $q_s$ is a function of both $\Omega$ and $\beta$ \HA{and shall dictate the new definition of BZ}. Following the parametrization in Eq.~(\ref{eq:mag_phase}), the dispersion relation in (\ref{eq:non-dim_driven_wave}) \HA{is recast} to:
\begin{equation}
    \cos(q-q_s) = \frac{2(1-\beta^2)-\Omega^2}{2\sqrt{\beta^2 \Omega^2 + (1-\beta^2)^2}}
    \label{eq:phase_shift}
\end{equation}
and the driven-wave formulation of the dispersion relation \HA{is} obtained by solving for $q$:
\begin{equation}
    q = \pm \cos^{-1} \left(\frac{2(1-\beta^2)-\Omega^2}{2\sqrt{\beta^2 \Omega^2 + (1-\beta^2)^2}} \right) + q_s
    \label{eq:q_driven}
\end{equation}
\HA{For $\beta \neq 0$, Eq.~(\ref{eq:q_driven}) output is no longer bounded by the traditional BZ boundaries $q\in[-\pi,\pi]$, evincing the shifted nature of BZ due to the phase shift $q_s$.~Owing to the group velocity vanishing at BZ boundaries for periodic media \cite{martinez2017standing}, WML's group velocity must be first derived to precisely pinpoint its BZ boundaries.~Then, the BZ shift shall be obtained by evaluating $q_s$ at the frequency corresponding to a zero group velocity.~\HA{To do so, consider the (non-dimensional) group velocity $c_g$ for WML, derived by taking the derivative of Eq.~(\ref{eq:disp_branch}) with respect to the wavenumber $q$ (See Supplementary Note 1 for more discussion on group velocity and phase shift):
\begin{equation}
    c_g(q) = \frac{\partial \Omega}{\partial q} = \frac{\sin(q) \left(1 - 2\beta^2 \sin^2\left(\frac{q}{2}\right) \right)}{2\left|\sin\left(\frac{q}{2}\right)\right|\sqrt{1 -\beta^2 \sin^2\left(\frac{q}{2}\right)}} -\beta \cos(q) 
    \label{eq:c_g}
\end{equation}
By equating the group velocity in Eq.~(\ref{eq:c_g}) to zero, the following wavenumber roots are found:
\begin{equation}
    q_{\text{max}} = 4\tan^{-1} \left( \beta \pm \sqrt{1+\beta^2} \right)
    \label{eq:wavenumber_at_max_OM}
\end{equation}
Substituting Eq.~(\ref{eq:wavenumber_at_max_OM}) into Eq.~(\ref{eq:disp_branch}) reveals that the frequency at a zero group velocity is $\Omega_{\text{max}} = 2$, which is constant, independent of $\beta$, and equivalent to the cutoff frequency of a reciprocal ML. Finally, the BZ shift, denoted as $\phi$ and constitutes the key result in this study, is derived by evaluating $q_s$ at $\Omega_{\text{max}} = 2$:}
\begin{equation}
    \phi = 
    \tan^{-1} \left(\frac{2\beta}{1-\beta^2} \right)
\label{eq:BZ_Shift}
\end{equation}
}

Equations~(\ref{eq:q_driven})~and~(\ref{eq:BZ_Shift}) \HA{uncover} that the BZ in its entirety is \textit{shifted} and it is within the range $q \in [-\pi+\phi,\pi+\phi]$, with its width remaining \textit{constant} at $2\pi$ in analogy to a reciprocal ML.~\HA{Depending on the sign of $\beta$ (and subsequently $\phi$)}, the BZ moves forward or backward, \HA{as seen in the middle} row of Figure~\ref{fig:Dispersion}.~Not only that non-reciprocity shifts the BZ, but it also forces the range of the wavenumber corresponding to the forward-going and backward-going waves to be \textit{unequal}.~\HA{To maintain a constant BZ width of $2\pi$, the amount of shrinkage (or extension)} in the forward-going wave wavenumber range (i.e., $q \in [0,\pi+\phi]$) is compensated by a larger (smaller) wavenumber range of the backward-going waves (i.e., $q \in [-\pi+\phi,0]$), \HA{depending on the sign of $\phi$}.~This shift $\phi$ disappears when the relative speed $\beta$ is zeroed out as expected, which is verifiable from Eq.~(\ref{eq:BZ_Shift}), \HA{and the traditional BZ range ($q \in [-\pi,\pi]$) is recovered}.

A few additional observations from the driven-wave dispersion in Figure~\ref{fig:Dispersion}:~(i) the attenuation at a frequency higher than the cutoff frequency $\Omega_{\text{max}} =2$ is smaller in WMLs compared to a reciprocal ML.~(ii) The real component of the wavenumber with $\Omega_{\text{max}}>2$ \HA{in WMLs} does not have a constant value and varies as the frequency increases, unlike its reciprocal counterpart with a constant real wavenumber of $\pm \pi$ within the attenuation zone ($\Omega_{\text{max}} >2$). (iii) The BZ defined in MLs with $\beta = 0$ is symmetric about its center, allowing for an irreducible BZ to be defined within the range $q \in [0,\pi]$. As \HA{WMLs have} asymmetric dispersion branches, reducing the first BZ may be elusive.~A zero wavenumber, however, remains the point that divides the forward- and backward-going waves for both lattices. \HA{Finally, the inequivalence of the forward and backward wavenumber ranges is numerically verified via a spatiotemporal fast-Fourier transform (FFT) of the time response of a finite WML (bottom row of Figure~\ref{fig:Dispersion}), exhibiting excellent agreement with the analytical results of the newly defined BZ definition in the middle row (See Supplementary Notes 2 for more details).}

\HA{
\section{Extension to two-dimensional lattices}

\begin{figure*}[]
     \centering
\includegraphics[width=0.95\textwidth]{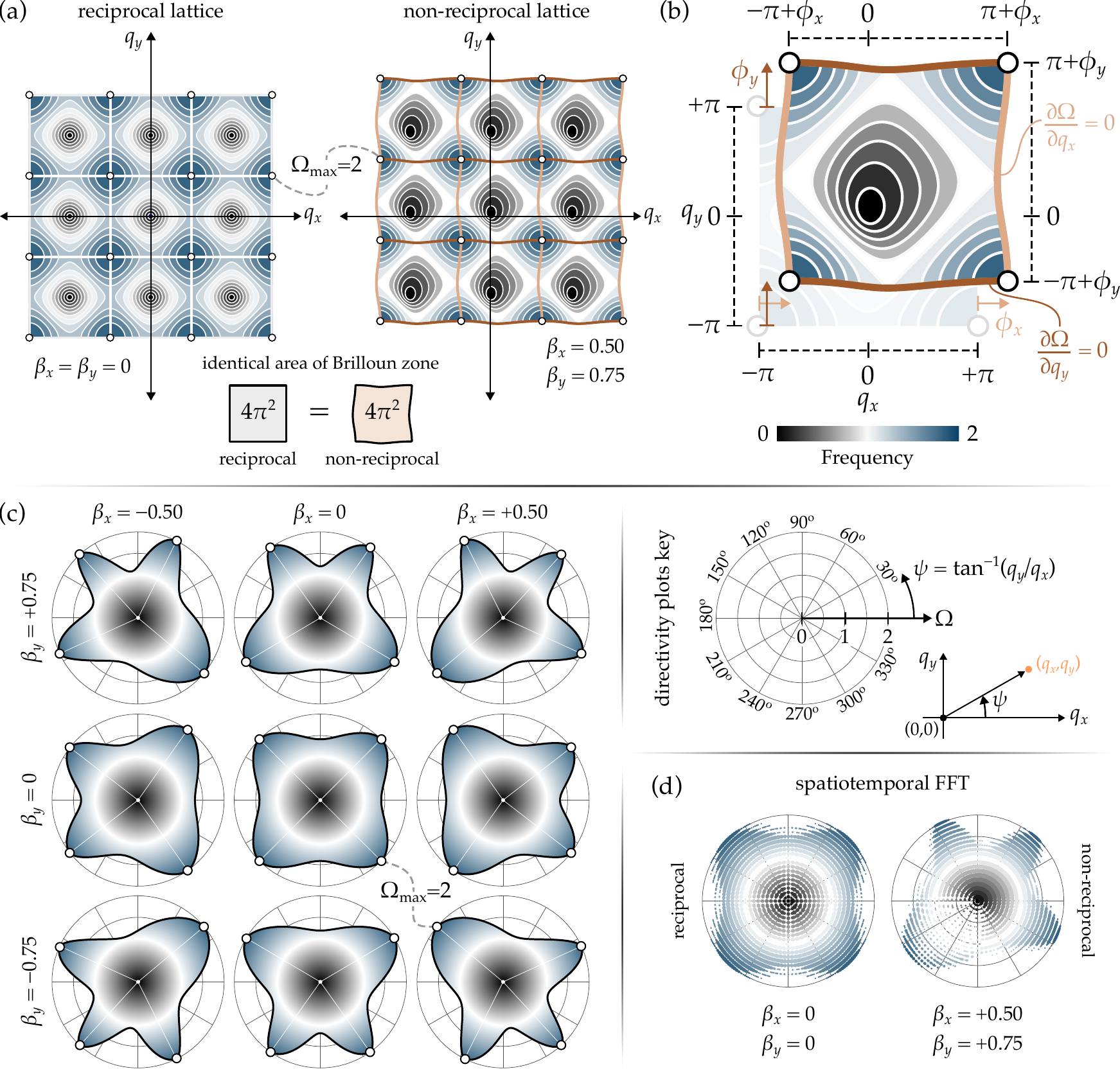}
\caption{\HA{(a) Dispersion contours of a reciprocal square ML ($\beta_x = \beta_y = 0$) and a non-reciprocal square WML ($\beta_x = +0.5$ and $\beta_y = +0.75$), highlighting the definition of BZ in each case. For the WML case, the BZ area, while irregular in shape, is identical to that of its reciprocal counterpart. (b) A close-up of BZ for square WMLs, illustrating the shift in the $x$ and $y$ propagation directions (i.e., $\phi_x$ and $\phi_y$, respectively), as well as the vanishing group velocity in the $x$-direction ($y$-direction) for the \textit{deformed} vertical (horizontal) BZ boundaries. (c) Directivity plots for a variety of combinations of $\beta_x$ and $\beta_y$, using the proposed definition of BZ for square WMLs. The depicted white lines correspond to the directions at which the cutoff frequency occurs. (d) Numerically-constructed directivity plots for $\beta_x = \beta_y = 0$ (reciprocal case) and $\beta_x = +0.5$ and $\beta_y = +0.75$ (non-reciprocal case), achieved via spatiotemporal FFT, showing excellent agreement to their theoretical counterpart in sub-figure~(c).}}
     \label{fig:2d-extension}
\end{figure*}

Next, the analysis for the one-dimensional WML is extended to a two-dimensional counterpart,~\HBA{using an elastic membrane with velocity modulations $v_x$ and $v_y$ in the $x$ and $y$ directions, respectively, and a proposed govering equation of:}
\begin{equation}
    \frac{\partial^2 u}{\partial t^2} + v_x \frac{\partial^2 u}{\partial x \partial t} + v_y \frac{\partial^2 u}{\partial y \partial t} + \frac{1}{2}\left( \left(v^2_x- c^2 \right) \frac{\partial^2 u}{\partial x^2} + \left(v^2_y- c^2 \right) \frac{\partial^2 u}{\partial y^2} \right) = 0
    \label{eq:EOM-moving-rod-2d}
\end{equation}
where $u(x,y,t)$ symbolizes the membrane's transverse displacement. \HBA{After discretizing Eq.~(\ref{eq:EOM-moving-rod-2d}) as a square grid, it can be shown that the dispersion relation for the resulting square WML is (See Supplementary Note 3):}
\begin{equation}
      \Omega^2 + \Omega \left(\beta_x \sin(q_x) + \beta_y \sin(q_y) \right) - 2\left((1-\beta_x^2)\sin^2 \left( \frac{q_x}{2} \right) + (1-\beta_y^2) \sin^2 \left( \frac{q_y}{2} \right) \right) = 0
    \label{eq:non-dim_disp_rel_2d_1}
\end{equation}
where $q_{x,y}$ and $\beta_{x,y} = v_{x,y}/c$ are the wavenumber and relative velocity in the $x$ (or $y$) direction, respectively.~\HA{Note that $\beta_{x,y} \in [-1,1]$ and the dispersion relation of a traditional square ML is recovered if $\beta_x = \beta_y = 0$, as expected.~The dispersion surface $\Omega$ of the square WML in Eq.~(\ref{eq:non-dim_disp_rel_2d_1}) is found by solving the quadratic equation for values of $q_x$ and $q_y$ within a newly defined (and shifted) BZ}. To reveal such a BZ, an identical procedure to the one used in deriving Eq.~(\ref{eq:phase_shift}) is followed here, yielding an alternative form of Eq.~(\ref{eq:non-dim_disp_rel_2d_1}):
\begin{equation}
      \Omega^2 - (2-\beta_x^2-\beta_y^2) + |z_{x}| \cos(q_x - q_s^x) + |z_{y}| \cos(q_y - q_s^y) = 0
    \label{eq:non-dim_disp_rel_2d}
\end{equation}
where
\begin{equation}
    |z_{x,y}| = \sqrt{\beta_{x,y}^2 \Omega^2 + (1-\beta_{x,y}^2)^2}; \ \ \ \ \ q_s^{x,y} = \tan^{-1}\left(\frac{\Omega \beta_{x,y}}{1-\beta_{x,y}^2} \right)
    \label{eq:mag_phase-2d}
\end{equation}
Numerically, it can be shown that the dispersion relation of the square WML has a constant cutoff frequency at $\Omega_\text{max} = 2$, regardless of $\beta_x$ and $\beta_y$ (See Supplementary Figure~\ref{fig:constant_om_BZ}(a)).~Analogous to Eq.~(\ref{eq:BZ_Shift}), and after plugging in $\Omega_\text{max} = 2$ back into $q_s^{x,y}$ in Eq.~(\ref{eq:mag_phase-2d}), it is concluded that the two-dimensional BZ corners are shifted with $x$ and $y$ direction shifts of:
\begin{equation}
    \phi_{x,y} = \tan^{-1} \left( \frac{2\beta_{x,y}}{1-\beta_{x,y}^2} \right)
\end{equation}

In a reciprocal square ML, the vertical (horizontal) boundaries of its perfectly square BZ correspond to a vanishing group velocity in the $x$-direction ($y$-direction) only. In extension, the same condition shall be imposed on the non-reciprocal WML case to find the horizontal and vertical boundaries of the shifted BZ. By doing so, it is found that the BZ for WML is no longer a square and has irregularly shaped boundaries.~Yet, its area remains a \textit{constant} value of $4 \pi^2$, \textit{identical} to its reciprocal case, regardless of the magnitude of $\beta_x$ and $\beta_y$ (See Figure~\ref{fig:2d-extension}(a,b) and Supplementary Figure~\ref{fig:constant_om_BZ}(b)). 

Another way to depict the two-dimensional dispersion relation is by using directivity plots and reducing the two variables $q_x$ and $q_y$ into a single (direction) angle equal to $\psi = \tan^{-1}(q_y/q_x)$. In the reciprocal case, all four quadrants of the angle $\psi$ show identical profiles of the dispersion surface, and the same cannot be said when $\beta_x \neq 0$ and/or $\beta_y \neq 0$ (Figure~\ref{fig:2d-extension}(c)).~Complete dispersion surfaces in the directivity plot are generated if the wavenumber values within the non-reciprocal BZ boundaries (shown in Figure~\ref{fig:2d-extension}(b)) are substituted to the dispersion relation, which is not the case if the traditional BZ values of the wavenumbers $q_x$ and $q_y$ are used instead (See Figure~\ref{fig:2d-extension}(c) and Supplementary Figure~\ref{fig:2D_Contours}). Finally, a spatiotemporal FFT of the time response of a square WML of finite size is shown in Figure~\ref{fig:2d-extension}(d) for $\beta_x = \beta_y = 0$ (reciprocal case) and $\beta_x = 0.5$ and $\beta_y = 0.75$ (non-reciprocal case). The FFT contours exhibit a close agreement with their analytical counterparts in Figure~\ref{fig:2d-extension}(c), substantiating the approach's validity and the completeness of the newly defined BZ (See Supplementary Note 2 for more details).}

\section{Concluding remarks}
In summary, this study \HBA{demonstrates} that the Brillouin zone (BZ) remains constant in width\HA{/area} with non-reciprocity in a class of \HA{Willis} monatomic lattices (WMLs), yet its boundaries are shifted in a precisely quantifiable amount.~Synthesized from a modified wave equation of longitudinal waves in moving elastic rod~\cite{attarzadeh2018elastic}, the {one-dimensional WMLs} is studied analytically and its driven-wave dispersion relation is proven vital for the quantification of the BZ shift $\phi$.~It is shown that the proposed theory perfectly agrees with the numerical simulation, {validating BZ shifting, its constant width, and inequivalence of wavenumber ranges occupying the forward-going and backward going waves. It is also established that an increase (decrease) in the forward-going wavenumber range is compensated by a shrinkage (enlargement) in the backward-going one.~An extension to a square WML is also established, demonstrating the shifted nature of the BZ and its area being constant and unaffected by the degree of non-reciprocity, dictated by the velocity modulation in the $x$ and $y$ direction. Additionally, it is shown that the two-dimensional BZ is no longer a square, and its deformed vertical (horizontal) boundaries are found by setting the group velocity in the $x$ ($y$) propagation direction to zero}.~The established \HBA{clarification on the definition of BZs} is envisioned to be a stepping stone for further investigations in BZ quantification for different types of modulations for wave non-reciprocity, \HA{as well as for periodic media with multi-mode dispersion relations.}

\section*{References}
\begin{multicols}{2}
\footnotesize
\printbibliography[heading=none]
\end{multicols}
\break

\noindent 
\large{\underline{\smash{Supplementary Material}}}
\setcounter{figure}{0}
\renewcommand{\thefigure}{S\arabic{figure}}
\setcounter{equation}{0}
\renewcommand{\theequation}{S\arabic{equation}}
\section*{\large{Note 1.~Group velocity and phase shift in one-dimensional Willis monatomic lattice}}
\begin{figure*}[h]
     \centering
\includegraphics[width=0.95\textwidth]{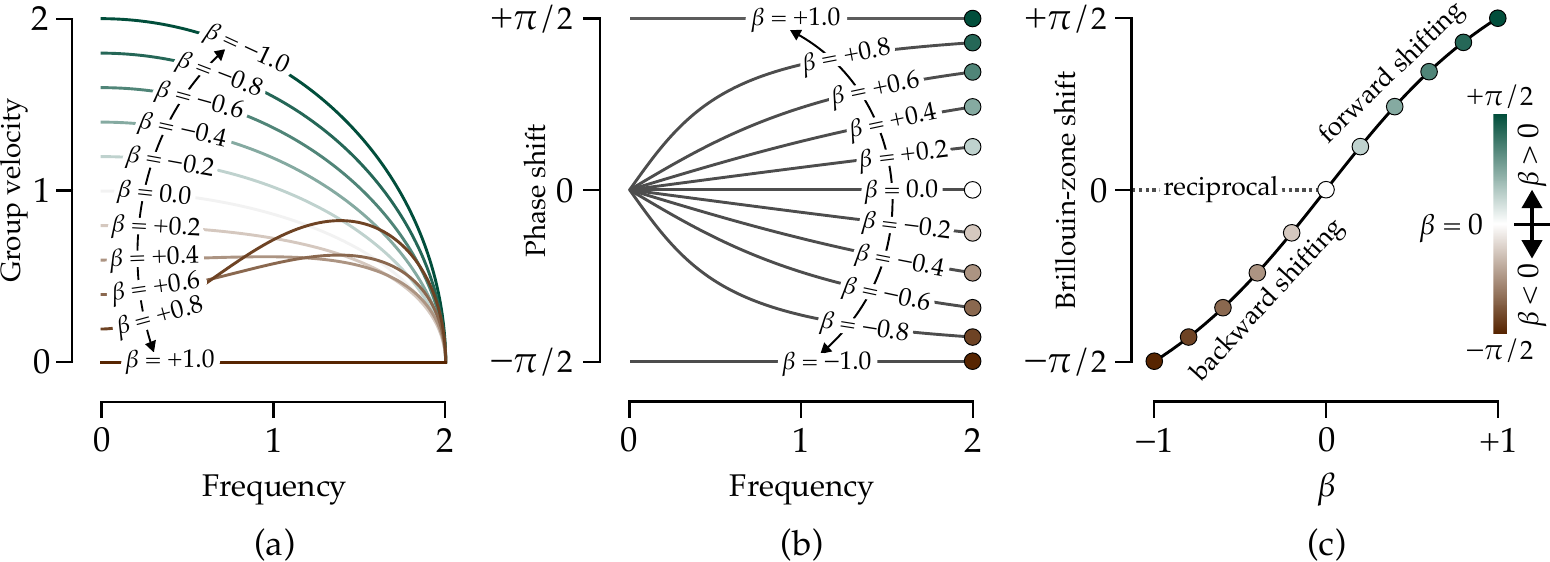}

\caption{\HA{(a) Group velocity $c_g$ for the forward-going waves $q \in [0,\pi+\phi]$}, \HA{(b)} Phase shift $q_s$ and \HA{(c)} the Brillouin zone shift $\phi$ as a function of the relative velocity $\beta$. Note that the output of the \HA{group velocity $c_g$ and} phase shift $q_s$ also depends on the excitation frequency $\Omega$. The shift of the Brillouin zone $\phi$ signifies the values of phase shift $q_s$ at a frequency of $\Omega_\text{max} = 2$, and, thus is only a function of $\beta$.}
     \label{fig:phase}
\end{figure*}

\normalsize
For completeness, \HA{the group velocity $c_g$}, the phase shift $q_s$, and the Brillouin zone shift (corresponds to $q_s$ at the cutoff frequency of $\Omega_{\text{max}} =2$) are shown in Figure~\ref{fig:phase} \HA{for different values of $\beta$}. \HA{As proven earlier, the group velocity (shown here for the forward-going waves $q \in [0,\pi+\phi]$) vanishes at the cutoff frequency $\Omega_\text{max} = 2$ (corresponding to $q = \pi+\phi$) regardless of the magnitude of the relative velocity $\beta$. Worthy of note is that the initial slope of the dispersion branch (i.e., the group velocity at $\Omega = 0$) is $c_0 = 1-\beta$ ($c_0 = -1-\beta$) for dispersion branches pertaining to forward-going (backward-going) waves. For the phase shift $q_s$,} it is observed that \HA{its} range is within $[-\pi/2, \pi/2]$, corresponding to a range of relative velocity of $\beta \in [-1, 1]$ and frequency $\Omega \in [0,2]$. \HA{Being frequency-dependent is another clear characteristic of phase shift $q_s$ as seen from the series of curves in Figure~\ref{fig:phase} with $\beta \neq 0$, where $q_s$ appears to increase} as the frequency increases.~The Brillouin zone shift $\phi$, on the other hand, does not depend on~$\Omega$ since it corresponds to the phase shift $q_s$ at the \textit{constant} cutoff frequency $\Omega_{\text{max}} =2$. 

\section*{\large{Note 2.~Numerical validation for the dispersion relations of Willis monatomic lattices}}

\begin{figure*}[]
     \centering
\includegraphics[]{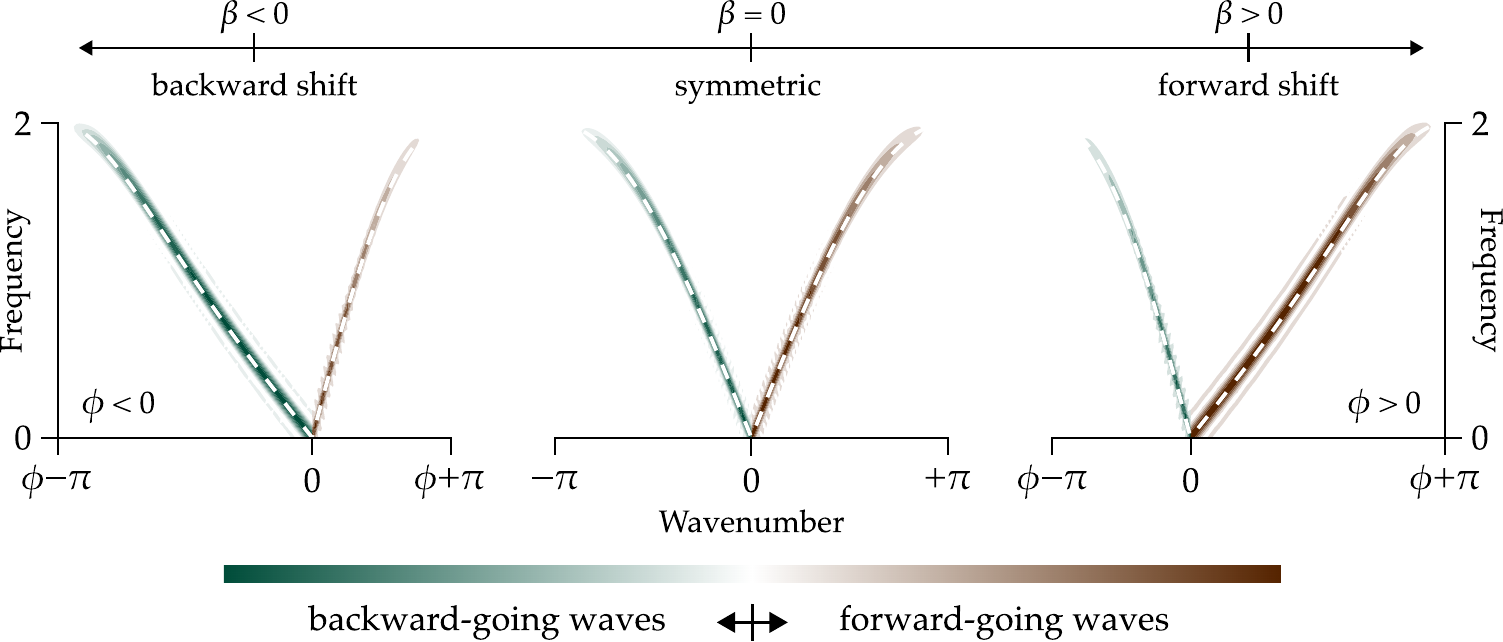}
\caption{Spatiotemporal FFT for the impulse response of the one-dimensional monatomic lattice, applied on the left and right ends for quantifying forward- and backward-going wavenumbers, respectively.~Forward (backward) shift is observed with positive (negative) relative velocity $\beta$, which also correspond to a positive (negative) phase shift value of $\phi$.~A zero value of $\beta$ returns the conventional dispersion relation of a reciprocal monatomic lattice with un-shifted BZ and symmetric dispersion branches. Results for non-reciprocal cases are for $|\beta | =0.5$, as in Figure~2 in the main manuscript.}
     \label{fig:2DFFT}
\end{figure*}

Numerical simulations are performed using a finite lattice of fixed boundaries and sufficient size to support the theoretical results and further validate the shift of the boundaries of the Brillouin zone (BZ). For the one-dimensional case, and to accurately quantifying the shift in BZ, the wavenumbers corresponding the forward-going (backward going) waves must be excited separately. Therefore, an impulse excitation is injected to a finite lattice of 151 masses at its left (right) end and the simulation time is chosen such that the impulse does not reach the other end and reflects (See Supplementary Videos S1-S6 for the finite lattice response for $\beta =\pm0.5$ and $\beta~=~0$).~A spatiotemporal fast Fourier transform (FFT) is then performed to construct the numerical dispersion diagram from the impulse response of the lattice.~The dispersion branch for the forward-going (backward-going) waves are constructed from the numerical response of the lattice for the impulse excitation on the left (right) end, which guarantees that the wavenumbers corresponding to forward-going and backward-going waves can be quantified separately.~As seen in Figure~\ref{fig:2DFFT}, the wavenumber spanning the forward-going (backward-going) waves is smaller (larger) than the conventional range in a reciprocal media when $\beta < 0$ ($\beta > 0$), in agreement with the theoretical results (shown as white dashed-lines). It is now evident that the numerical simulation corroborates that BZ is no longer in the conventional range and its shift is precisely quantifiable based on the phase shift~$\phi$. 

For the square lattices, an initial displacement was applied on the middle of a $40 \times 40$ lattice to excite all modes of the structure, and the simulation is terminated before reflections from the boundaries occur. Performing spatiotemporal FFT and projecting its amplitude on a directivity plot, the result is depicted in Figure~\ref{fig:2d-extension}(d) in the main manuscript. The wavenumbers used for this procedure are those for the newly proposed BZ, resulting in dispersion plots that are in excellent agreement with the analytically obtained one in Figure~\ref{fig:2d-extension}(c) in the main manuscript.~For better results visualization, FFT's magnitude of 10\% or smaller was filtered.

\section*{\large{Note 3.~Equation of motion and dispersion relation of square Willis monatomic lattices}}

\normalsize
The proposed equation of motion for a two-dimensional, non-reciprocal continuous elastic medium is:
\begin{equation}
    \frac{\partial^2 u}{\partial t^2} + v_x \frac{\partial^2 u}{\partial x \partial t} + v_y \frac{\partial^2 u}{\partial y \partial t} + \frac{1}{2}\left( \left(v^2_x- c^2 \right) \frac{\partial^2 u}{\partial x^2} + \left(v^2_y- c^2 \right) \frac{\partial^2 u}{\partial y^2} \right) = 0
    \label{eq:EOM-moving-rod-2d_2}
\end{equation}
which correspond to the following mass, gyral, and stiffness operators of Eq.~(\ref{eq:EOM_operators}) in the main manuscript:
\begin{equation}
    M = 1, \ \ \ \ G = v_x \frac{\partial }{\partial x} + v_y \frac{\partial }{\partial y}, \ \ \ \ K =  \frac{1}{2}\left(v_x^2\frac{\partial^2 }{\partial x^2} + v_y^2 \frac{\partial^2 }{\partial y^2} \right) - \frac{c^2}{2} \left( \frac{\partial^2}{\partial x^2} + \frac{\partial^2 }{\partial y^2} \right)
    \label{eq:operators_2D}
\end{equation}
Using central finite difference on the spatial domains, the discretization for the mixed derivatives, as well as the $x$ and $y$ second derivatives are as follows:
\begin{subequations}
   \begin{equation}
    \frac{\partial^2 u}{ \partial x \partial t} = \frac{\dot{u}_{i+1,j} - \dot{u}_{i-1,j}}{2a_x}
\end{equation}
\begin{equation}
    \frac{\partial^2 u}{ \partial y \partial t} = \frac{\dot{u}_{i,j+1} - \dot{u}_{i,j-1}}{2a_y}
\end{equation}
\begin{equation}
    \frac{\partial^2 u}{ \partial x^2} = \frac{u_{i+1,j} - 2u_{i,j} + u_{i-1,j}}{a_x^2}
\end{equation}
\begin{equation}
    \frac{\partial^2 u}{ \partial y^2} = \frac{u_{i,j+1} - 2u_{i,j} + u_{i,j-1}}{a_y^2}
\end{equation} 
\label{eq:central_differencing}
\end{subequations}
where $[ \dot \ ]$ represents the derivative in time, $t$. Assuming a square grid, i.e., $a_x = a_y = a$, and using the discretization schemes shown in Eq.~(\ref{eq:central_differencing}), one can show that the equation of motion for the $({i,j})^\text{th}$ unit cell is given by:
\begin{align}
\begin{split}
    \ddot{u}_{i,j} + \frac{v_x}{2a} \left( \dot{u}_{i+1,j} - \dot{u}_{i-1,j}\right) +  \frac{v_y}{2a} \left(\dot{u}_{i,j+1} -  \dot{u}_{i,j-1} \right)+ \frac{1}{2a^2} \left(c^2 - v_x^2 \right)\left(2u_{i,j} - {u}_{i+1,j}-u_{i-1,j} \right) \\
     + \frac{1}{2a^2} \left(c^2 -v_y^2 \right)\left(2u_{i,j} - {u}_{i,j+1}-u_{i,j-1} \right) = 0
    \end{split}
\label{eq:EOM_moving_monatomic_lattice_2d_1}
\end{align}

\begin{figure*}[]
     \centering
\includegraphics[]{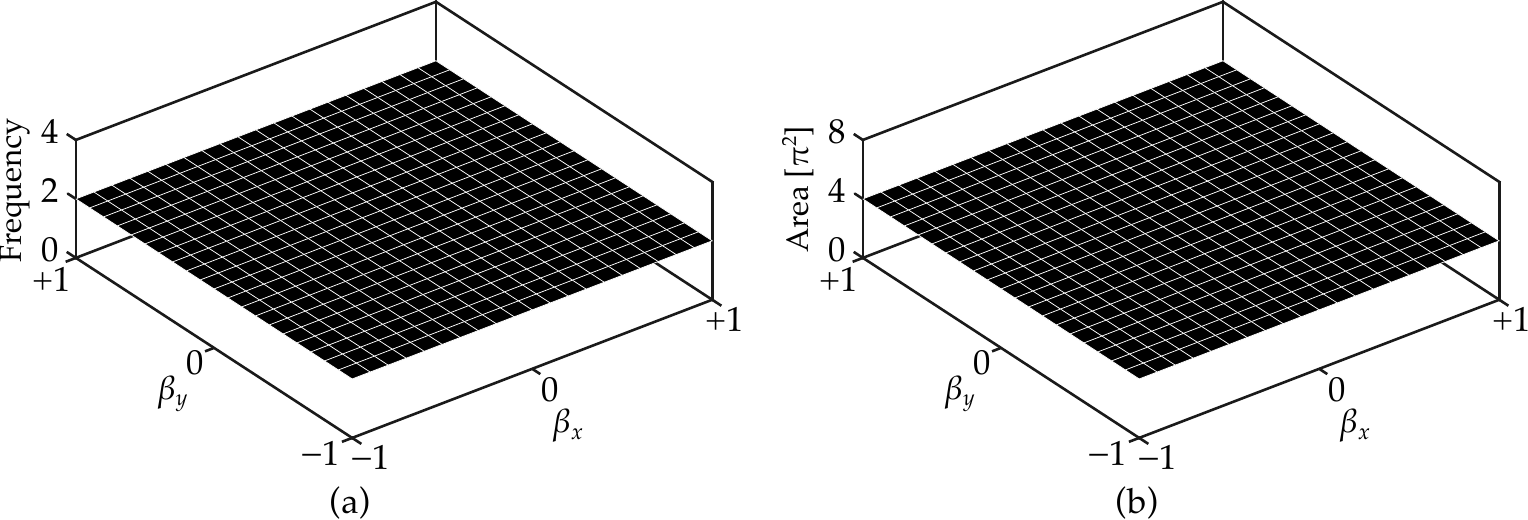}
\caption{Numerical calculation of (a) the cutoff frequency and (b) the Brillouin zone area for the square Willis monatomic lattice over the entire range of $\beta_x$ and $\beta_y$, showing their independence on these two parameters.}
     \label{fig:constant_om_BZ}
\end{figure*}

\begin{figure*}[h]
     \centering
\includegraphics[]{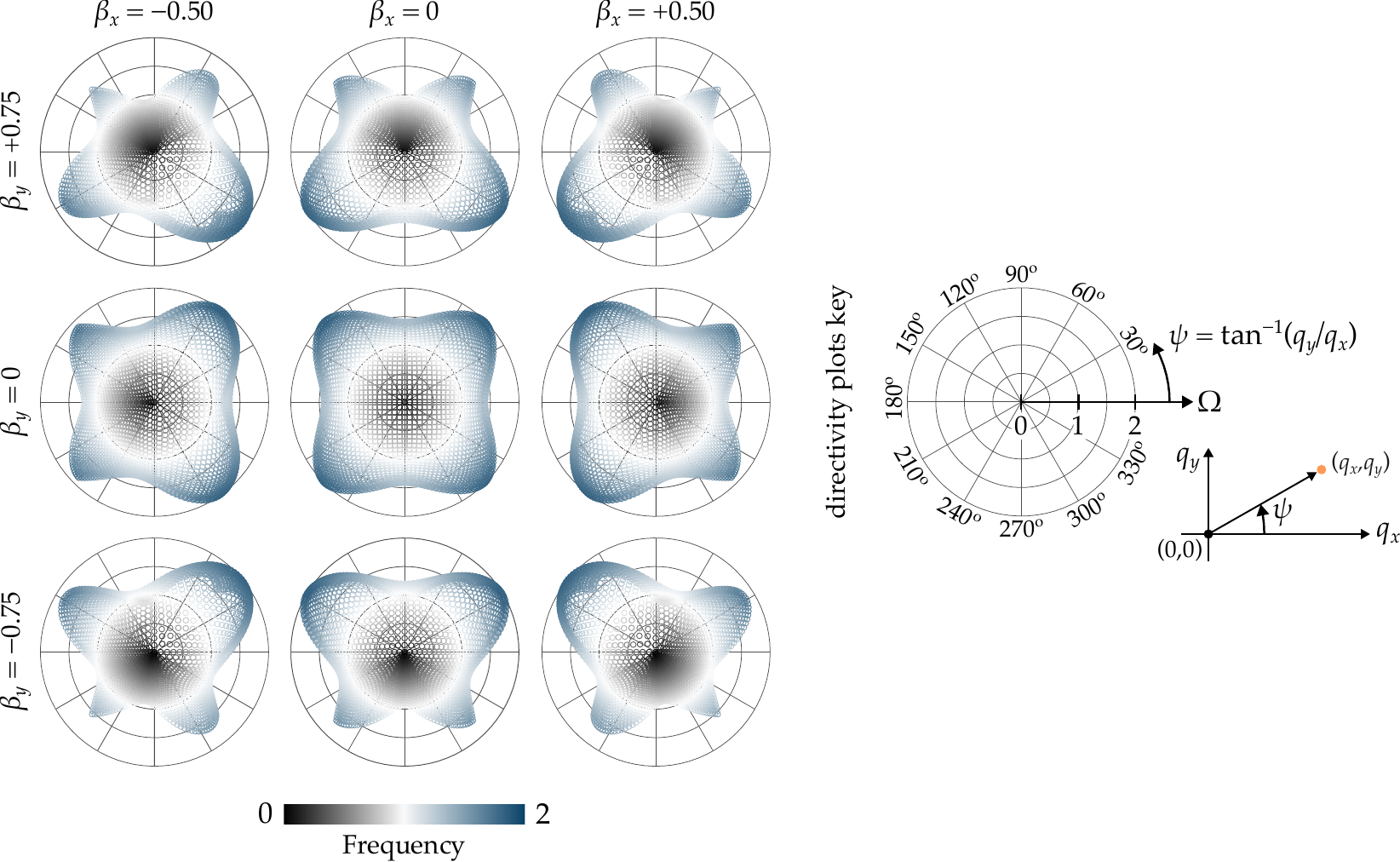}
\caption{Contours of dispersion relations for square Willis monatomic lattices with a variety of combinations of $\beta_x$ and $\beta_y$, presented via directivity plots, using the traditional definition of BZ for square lattices. As can be seen, the dispersion plots are not complete, in sharp contrast to the contours plotted using the newly proposed BZ definition shown in Figure~\ref{fig:2d-extension} in the main manuscript.}
     \label{fig:2D_Contours}
\end{figure*}

Knowing that $\omega_0 = c/a$ and using an equivalent mass of $m$ per unit cell, the equation of motion~(\ref{eq:EOM_moving_monatomic_lattice_2d_1}) can be re-written in a form incorporating the mass $m$ and the stiffness parameter $k = m\omega^2_0$:
\begin{align}
\begin{split}
    m\ddot{u}_{i,j} + \frac{mv_x}{2a} \left( \dot{u}_{i+1,j} - \dot{u}_{i-1,j}\right) +  \frac{mv_y}{2a} \left(\dot{u}_{i,j+1} -  \dot{u}_{i,j-1} \right)
    + \frac{1}{2} \left(k -\frac{m v_x^2}{{a^2}} \right)\left(2u_{i,j} - {u}_{i+1,j}-u_{i-1,j} \right) \\
    + \frac{1}{2} \left(k -\frac{m v_y^2}{{a^2}} \right)\left(2u_{i,j} - {u}_{i,j+1}-u_{i,j-1} \right) = 0
    \end{split}
\label{eq:EOM_moving_monatomic_lattice_2d_2}
\end{align}
Upon the substitution of the Bloch theorem and making use of $\omega_0 = c/a$ and $v_{x,y} = c \beta_{x,y}$, the equation of motion~(\ref{eq:EOM_moving_monatomic_lattice_2d_1}) boils down to:
\begin{equation}
 \ddot{u}_{i,j} + \mathbf{i} \omega_0 \left(\beta_x \sin(q_x) + \beta_y \sin(q_y) \right) \dot{u}_{i,j} 
    + 2 \omega_0^2 \left( \left(1 - \beta^2_x \right)\sin^2\left( \frac{q_x}{2}\right)+ \left(1 - \beta^2_y \right)\sin^2\left( \frac{q_y}{2}\right) \right){u}_{i,j}  = 0
\label{eq:EOM_moving_monatomic_lattice_2d_3}
\end{equation}
Assuming harmonic motion and normalizing the excitation frequency using $\omega_0$, the following non-dimensional dispersion relation for the square Willis monatomic lattice is obtained, which is identical to Eq.~(\ref{eq:non-dim_disp_rel_2d_1}) in the main manuscript.~The BZ corresponding to Eq.~(\ref{eq:non-dim_disp_rel_2d_1}) is no longer a square as in reciprocal square MLs. It is, instead, of irregular deformed shape and its boundaries are defined based on setting the group velocity of the $x$-direction ($y$-direction) waves to zero to find the vertical (horizontal) boundaries, as proven in the main manuscript. Finally, it is important to re-iterate that: 
\begin{itemize}
    \item The dispersion relation has a constant cutoff frequency of $\Omega_\text{max} = 2$, regardless of the magnitude of $\beta_x$ and $\beta_y$, as shown in Figure~\ref{fig:constant_om_BZ}(a).
    \item The BZ of square Willis monatomic lattices has an identical area to that of their reciprocal counterpart, regardless of the magnitude of $\beta_x$ and $\beta_y$, which is numerically shown in Figure~\ref{fig:constant_om_BZ}(b).~The change in the BZ shape with different combinations of $\beta_x$ and $\beta_y$ is shown in Supplementary Movie S7.
    \item Using the traditional range of BZ for square lattices, i.e., $q_{x} \in [-\pi,\pi]$ and $q_{y} \in [-\pi,\pi]$, returns incomplete dispersion diagrams, which can be seen from the directivity plots shown in Figure~\ref{fig:2D_Contours} (See Figure~\ref{fig:2d-extension} in the main manuscript for the complete dispersion diagrams for the same combinations of $\beta_x$ and $\beta_y$).
\end{itemize}

If $q_y = 0$ (analogously $q_x = 0$) in Eq.~(\ref{eq:non-dim_disp_rel_2d_1}), the dispersion relation becomes very similar to that of the one-dimensional case presented in the main manuscript, i.e., Eq.~(\ref{eq:non-dim_disp_rel}).~The shift at zero group velocity for a one-dimensional version of the square lattice is not equal to $\phi_{x,y}$, indicating that the BZ boundaries, indeed, do not form a square anymore. This is proven from analyzing the wave propagation in the square lattice along a single direction only, i.e., one of the components of the wavenumber has to be zero. As such, the dispersion relation reduces to (shown here for $q_y = 0$, but identical procedure can be done for $q_x = 0$):
\begin{equation}
      \Omega^2 + \Omega \beta_x \sin(q_x) - 2\left((1-\beta_x^2)\sin^2 \left( \frac{q_x}{2} \right) \right) = 0
    \label{eq:non-dim_disp_rel_2d_reduced}
\end{equation}
Following identical procedure to that in the main manuscript to find the wavenumbers at which the cutoff frequency occurs (i.e., at a zero group velocity in the $x$-direction), the wavenumber solutions on BZ boundaries are:
\begin{subequations}
   \begin{equation}
    q_{x,1} = -\mathbf{sgn}[\beta_x] \cos^{-1}(\beta_x^2-1)
\end{equation}
\begin{equation}
    q_{x,2} = \mathbf{sgn}[\beta_x] \left(2\pi - \cos^{-1}(\beta_x^2-1) \right)
\end{equation}
\end{subequations}
Substituting this wavenumber back into the dispersion relation in Eq.~(\ref{eq:non-dim_disp_rel_2d_1}) and solving for the frequency $\Omega$, one can show that the cutoff frequency is a function of the parameter $\beta_x$ and occurs at:
\begin{equation}
    \Omega_{\text{max}} = \frac{1}{2} \left[\sqrt{(6-\beta_x^2)\beta_x^4 - 12 \beta_x^2 + 8} + \sqrt{\beta_x^4(2-\beta_X^2)} \right]
\end{equation}
which results in a phase shift $q_s^x$ that is smaller than $\phi_x$, showing that the BZ of square Willis monatomic lattices should be of irregular shape.

\end{document}